\documentclass[a4paper,11pt,showkeys,nofootinbib]{revtex4}
\usepackage{url,graphicx}
\usepackage{soul}
\usepackage{amsmath, amssymb, setspace}

\begin{document}
\title{Whole-body counter survey results 4 months after the Fukushima Daiichi NPP accident in Minamisoma City, Fukushima}
\author{Ryugo S.~Hayano PhD,%
\footnote{Correspondence should be addressed: R. Hayano, (\url{hayano@phys.s.u-tokyo.ac.jp}).}
  Yuni Watanabe}
\affiliation{Department of Physics, The University of  Tokyo, 7-3-1 Hongo, Bunkyo-ku, Tokyo 113-0033, Japan}
\author{Shuhei Nomura, MSc}
\affiliation{Department of Epidemiology and Biostatistics, School of Public Health, Imperial College London, Norfolk Place, London W2 1PG, United Kingdom}
\author{
Tsuyoshi Nemoto MD,
Masaharu Tsubokura MD,
Tatsuo Hanai,
Yuki Kumemoto,
Satoshi Kowata,
Tomoyoshi Oikawa MD PhD,
Yukio Kanazawa  MD PhD}
\affiliation{Department of Internal Medicine, Minamisoma Municipal General Hospital, Minamisoma, Fukushima 979-0033, Japan}

\begin{abstract}
Using the first WBC unit installed in Fukushima Prefecture after the accident, the radiocesium body burdens of 566 high-risk residents of Minamisoma city were measured in July 2011 at the Minamisoma Municipal General Hospital. 
The analysis of the data was challenging because this chair-type WBC unit did not have sufficient shielding against background gamma rays, and methods had to be developed to reliably compensate for the body-attenuated background radiation.
Fortunately, data for repeated tests of hospital staff members using both the chair-type and well-shielded FASTSCAN WBC units, installed in September 2011, were available, and could be used to check the validity of the analysis.
The CEDs of all subjects, estimated under the assumption of acute inhalation in March 2011, were found to be less than 1 mSv.

\end{abstract}
\maketitle

\section{Introduction}
\begin{figure}[b]
\includegraphics[width=0.6\columnwidth]{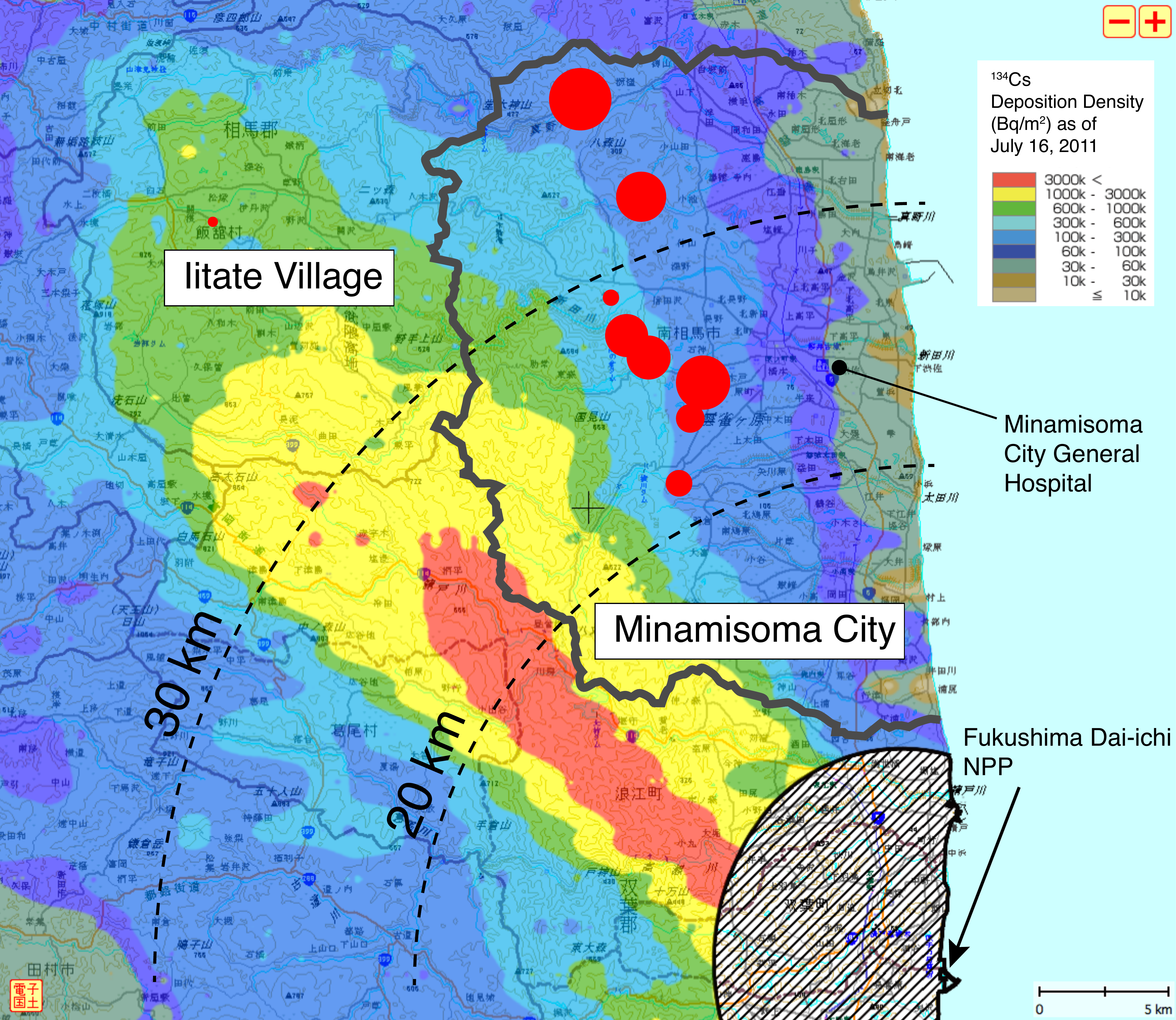}
\caption{\label{fig:map} 
The $^{134}$Cs deposition map as of July 16, 2011 (adapted from the Extension Site of Distribution Map of Radiation Dose, etc.,/Digital Japan \protect\url{http://ramap.jmc.or.jp/map/eng/map.html})  showing the vicinity of Minamisoma city. The black curve indicates the outline of Minamisoma city. The red circles indicate the geographical distribution of the residences of the ``Group-A'' subjects (see text), where the circles are proportional to the number of subjects living in each district. }\end{figure}

The severe accident involving the Fukushima Dai-ichi nuclear power plant (NPP), triggered by the Great East Japan Earthquake which took place on March 11, 2011, dispersed a large amount of radioactive materials into the environment~\cite{tanaka}. The Japanese government issued evacuation orders for people living within a 3~km radius of the Fukushima Dai-ichi NPP at 21:23 on March 11, within a 10~km radius at 5:44 on March 12, and within a 20~km radius at 18:25 on March 12. On March 15 at 11:00, the band between 20~km and 30~km radii was designated a ``shelter indoors'' zone. However, airborne monitoring surveys~\cite{mextairborne} carried out by the Japanese government (Fig.~\ref{fig:map}) show that the radioactive cesium deposition is not concentric; the most contaminated ``band'' extends some 40~km to the northwest of the Fukushima Dai-ichi NPP, into Iitate village.

The Minamisoma Municipal General Hospital (MMGH) is located 23~km north of Fukusihma Dai-ichi NPP, just 3~km outside the 20-km circle, in the ``shelter indoors'' zone. Because of this, as well as extensive tsunami damage to parts of the city and nearby areas, many difficulties were encountered by the hospital in providing basic medical services in the months following March, 2011, with particular difficulties in securing resources, manpower, medicine, and utilities. Nevertheless, since the western part of Minamisoma City borders the highly contaminated area, assessing the intake of radioactive materials in the early phase of the accident was considered by local medial staff to be essential and was given a high priority.

Thanks to the efforts of many people, a whole body counter (WBC) was installed at MMGH on  June 28, 2011. 
This was the first WBC unit installed in Fukushima Prefecture after the accident, providing local medical staff with the capability of measuring the internal contamination of residents. 
The unit was provided on loan for three months from the Ningyo-toge Environmental Engineering Center of the Japan Atomic Energy Agency, and had been installed in a bus-like vehicle which had been equipped for use as mobile radiation clinic in Tottori Prefecture. The vehicle was driven from there to Minamisoma, a distance of $\sim 1000$ km.
 It unfortunately arrived too late to be useful in detecting internal $^{131}$I, which has a half life $T_{1/2}$ of 8 days, but it could be used to detect $^{134}$Cs ($T_{1/2} = 2$ years) and $^{137}$Cs ($T_{1/2} =$ 30 years).
 
This WBC unit, manufactured by Anzai Medical Co., LTD in 2001, was a chair-type, model S.I.M., with a NaI detector (5-inch diameter $\times$ 3-inch thickness) mounted on the chair back~\footnote{The unit was also equipped with a smaller NaI detector for thyroid scan. This was however not used in the present study.}. 
The mobile WBC unit was parked in the parking lot of MMGH, where the radiation dose rate at 1.0m  height was $\sim 0.6 \mu$Sv/h; after the parking lot was washed with a high-pressure water cleaner, the dose rate decreased to $\sim 0.3\mu$Sv/h.

By placing iron and lead sheets underneath the mobile unit on June 30th, the background level at the WBC was further lowered to $\sim 0.18 \mu$Sv/h, but this was still too high for reliable measurements since this WBC did not have sufficient radiation shielding of its own.

\begin{figure}
\includegraphics[width=0.6\columnwidth]{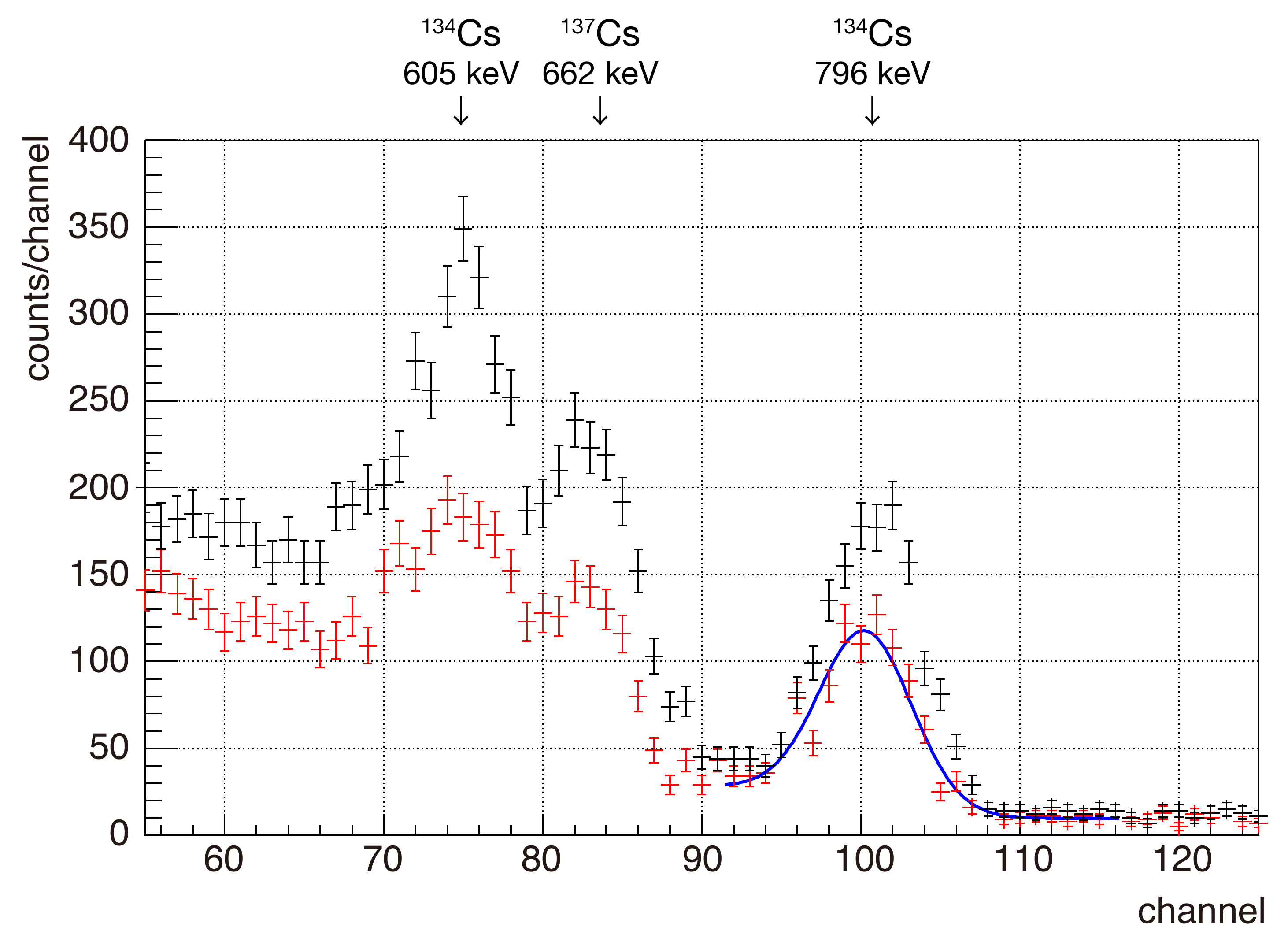}
\caption{\label{fig:spectra} Typical gamma-ray spectra measured at the Minamisoma Municipal General Hospital with a chair-type whole body counter in July, 2011. 
Black points represent the background spectrum, and red represents a subject with a moderate ($\sim 820$ Bq/body) internal contamination. The blue curve is the result of fitting the function defined in Eq.~A1 (Appendix~A) to the 796-keV peak.}
\end{figure}

This particular chair-type WBC unit, as well as most other chair-type units installed later in Fukushima Prefecture, was difficult to use with the required accuracy in the high-background radiation environment that existed after March 11, 2011. The major difficulty encountered is illustrated in Fig.~\ref{fig:spectra}: 
The spectrum shown in black is a background spectrum measured in July 2011. 
The right-most peak is caused by the 796 keV gamma-ray decay emission of $^{134}$Cs, and the double peaks at lower energy to the left result from emissions at 605 keV for $^{134}$Cs and 662 keV for $^{137}$Cs.
The spectrum drawn in red was obtained with a subject sitting on the chair. As can be seen, the radiocesium peak heights with the subject present are {\em lower} than those of the background alone. 
This is because the human body partially shields the environmental background radiation, the degree of shielding depending on the actual physical characteristics of the body, such as weight, height, girth, and so on.

In order to deduce the internal radiocesium contamination of human subjects in such a situation, it is therefore necessary to estimate the body-attenuated background spectrum (the black spectrum in Fig.~\ref{fig:spectra}  after being suppressed by a body-weight-and-height-dependent factor), and subtract it from the measured (red) spectrum.

This was by no means trivial, and very few applicable precedents exist. The required analysis, as well as subsequent testing and confirmation, was extremely time-consuming, and delayed the dissemination of this data.

\section{Methods} 
\subsection{Measurements}
At MMGH, the amount of radioactive cesium in the bodies of 566 citizens of Minamisoma was measured with the Anzai chair-type WBC between July 11, 2011 and July 29, 2011. 
The measurement time was 1 minute, shorter than fully desirable (with the consequence that the statistical accuracy was lower than for subsequent WBC measurements), but under the difficult conditions which prevailed during the summer of 2011, throughput, i.e. testing as many people as possible, was given precedence for this initial screening.

Priority was assigned to residents based on their suspected degree of exposure.  Highest priority was initially given to residents 
who lived in the most contaminated zone (Group A)  whose results we present in this paper. The demographic characteristics of the Group-A subjects are summarised in Table~\ref{tab:demography} (also see Fig.~\ref{fig:map}).
About 85\% of Group A residents evacuated from Minamisoma during March 2011 (about 10\% before March 13th, about 40\% before March 16th and about 82\% before March 21st), but have returned to Minamisoma in July 2011.  Detailed analyses of the internal contamination vs evacuation behaviors of the Group-A subjects will be presented in a forthcoming publication~\cite{nomura}.

In addition, beginning in July, 2011, data for repeated tests of MMGH staff members (Group X, 103 subjects) was obtained, which proved crucial for understanding the properties of the Anzai chair-type WBC.

Body height, weight, body fat percentage (BFP) as well as the evacuation behaviors of each subject were recorded.

This study was approved by the Ethics Committee of the University of Tokyo. 

\begin{table}
\caption{\label{tab:demography} The  demographic characteristics of the Group-A subjects.}
\begin{tabular}{ l r r }
\hline
Characteristics & Number & Percent \\ 
\hline
\multicolumn{3}{l}{Gender}\\ 
\hspace*{0.5cm}Male & 256 & 45.2 \\ 
\hspace*{0.5cm}Female & 310 & 54.8 \\ 
\multicolumn{3}{l}{Age at test} \\ 
\hspace*{0.5cm}16 -- 25 & 39 & 6.9 \\ 
\hspace*{0.5cm}26 -- 35 & 60 & 10.6 \\ 
\hspace*{0.5cm}36 -- 45 & 43 & 7.6 \\ 
\hspace*{0.5cm}46 -- 55 & 84 & 14.8 \\ 
\hspace*{0.5cm}56 -- 65 & 172 & 30.4 \\
\hspace*{0.5cm}66 -- 75 & 87 & 15.4 \\ 
\hspace*{0.5cm}76 -- 85 & 71 & 12.5 \\ 
\hspace*{0.5cm}86 -- 95 & 10 & 1.8 \\ 
\multicolumn{3}{l}{Address before the accident} \\ 
\hspace*{0.5cm}Minamicoma city &  &  \\ 
\hspace*{0.8cm}Kashima ward (outside 30 km zone) & 240 & 42.4 \\ 
\hspace*{0.8cm}Haramachi ward (inside 20 km zone) & 322 & 56.9 \\ 
\hspace*{0.5cm}Iitate village & 4 & 0.7 \\ 
\multicolumn{3}{l}{Address at test} \\ 
\hspace*{0.5cm}Outside Fukushima & 255 & 40.7 \\ 
\hspace*{0.5cm}Inside Fukushima & 244 & 44.1 \\ 
\hspace*{0.5cm}No evacuation & 84 & 15.2 \\ 
\hline
Total subjects & 566 & 100 \\ 
\hline
\end{tabular}
\end{table}

\subsection{Data analysis procedures}
We here describe the outline of the data analysis procedure. Technical details are described in Appendix {\ref{ap:spectra}-\ref{ap:burden}}.

\begin{figure}
\includegraphics[width=0.6\columnwidth]{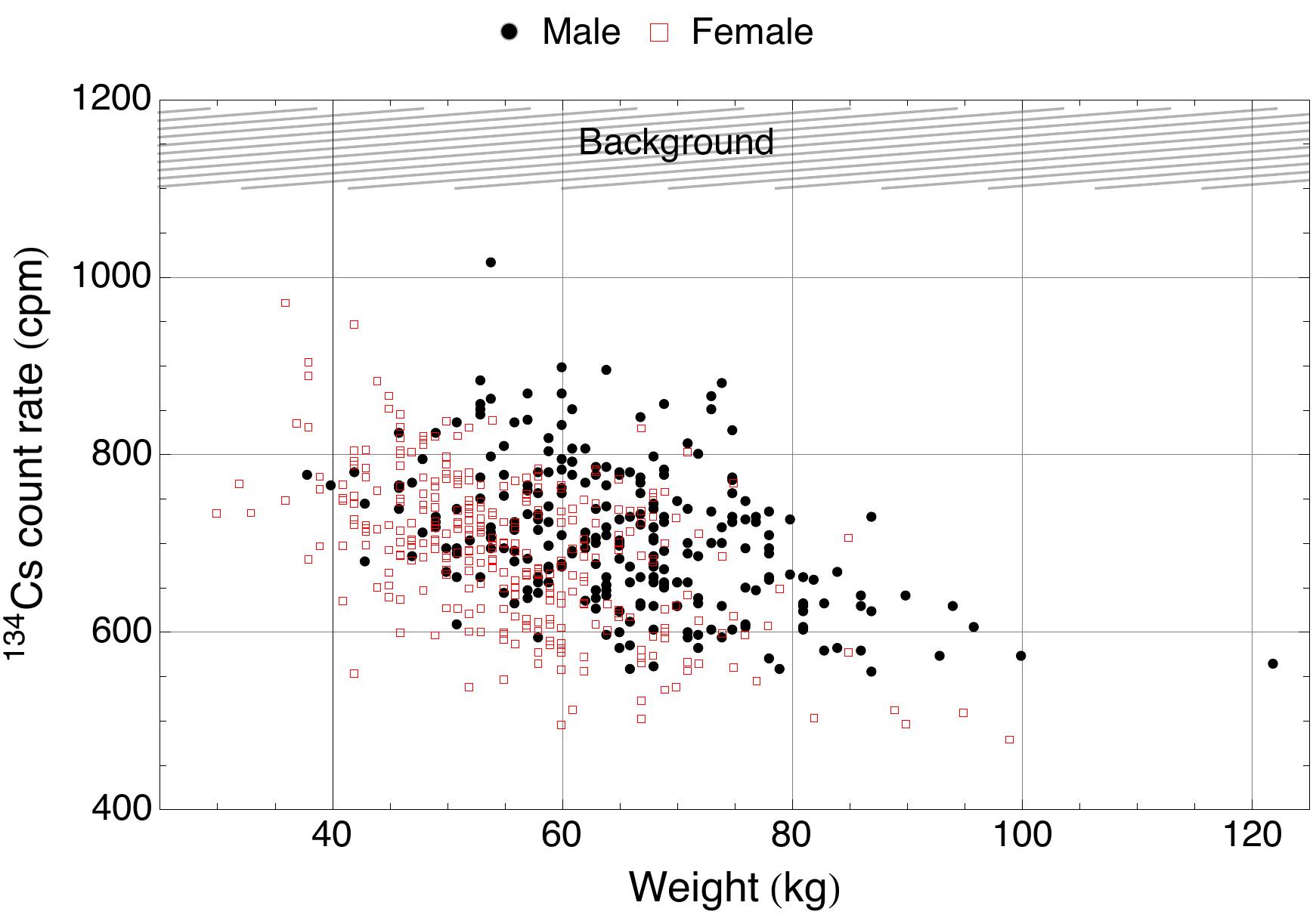}
\caption{\label{fig:weightvscs134} Body weight (kg) vs $^{134}$Cs count rate (cpm) for male subjects solid black circles) and for female subjects (red open squares). The hatched band indicates the background count rate.}
\end{figure}

\begin{figure*}
\includegraphics[width=\textwidth]{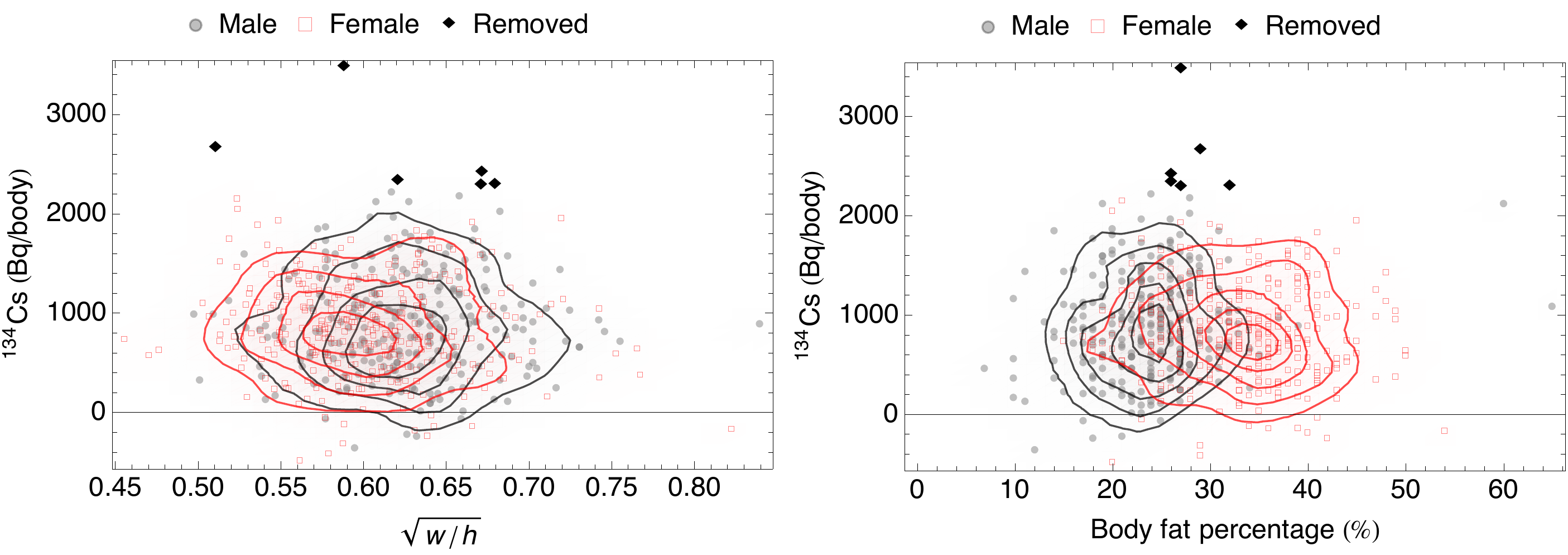}
\caption{\label{fig:contourswithbestparameters} The left panel shows the correlation between both thickness and $^{134}$Cs body burden, and the right panel shows body fat percentage versus $^{134}$Cs body burden. The data points for male subjects are shown as solid black circles, and females are shown as red open squares. The data points shown as black diamonds were not used in the determination of $\alpha$ and $\beta$ (see text).}
\end{figure*}

\begin{enumerate}
\item Spectra similar to the one shown in Fig.~\ref{fig:spectra} were obtained for each individual test.  For each, the count rate (counts per minute: cpm) of the $^{134}$Cs peak was obtained (see Appendix \ref{ap:spectra}).

\item The $^{134}$Cs count rate thus obtained however had a clear decreasing trend for heavier subjects, as shown in Fig.~\ref{fig:weightvscs134}, 
which can be attributed to the attenuation of the background by the subjects' bodies.
The background level measured without a subject (indicated by the gray hatched band) was about 1100-1200 cpm, higher by $\sim 200-500$ cpm than the count rate with a subject.

This body-attenuated background contribution was subtracted by using a phenomenological model, so that the dependence of the $^{134}$Cs count on the actual physical characteristics of the body, such as weight, height, and BFP was minimized. 
We first parameterized the background using the body thickness parameter $\sqrt{w/h}$, where $w$ represents body weight and $h$ represents body height (Eq.~\ref{eq:eq2}), but a more elaborate model including BFP (Eq.~\ref{eq:eq3}) was found necessary  (see Appendix \ref{ap:attenuate}). 

\item By using the detection efficiency coefficient (Appendix \ref{sec:efficiency}), $^{134}$Cs body burdens were estimated for the Group-A subjects (Appendix \ref{ap:burden}). The results are shown in Fig.~\ref{fig:contourswithbestparameters}, which depicts the correlation between the $^{134}$Cs body burden and: (left panel) the body thickness; (right panel) the body fat percentage. Neither of these distributions shows a clear dependence on $\sqrt{w/h}$  or on BFP, as intended.

%\item The validity of the fairly complex and indirect method of estimating internal $^{134}$Cs contamination outlined above was confirmed against an independent dataset of Group-X subjects (Appendix \ref{ap:verify}).

%\item Statistical and systematic errors were estimated, as described in Appendix \ref{ap:error}.
\end{enumerate}

\subsection{Verification of the present method}

The difficulty of the background radiation problem encountered in Minamisoma during these initial surveys led to a fairly complex and indirect method of estimating internal $^{134}$Cs contamination. It was essential to confirm the validity of these results against an independent dataset.

In September 2011, a better-shielded WBC (FASTSCAN Model 2251, Canberra Inc.) was installed at MMGH. 
The detection limit was 250 and 300 Bq/body for $^{134}$Cs and $^{137}$Cs each, using a 2-minute scan.
39 members of Group-X, comprised of 103 hospital staff in all, were scanned with both WBCs, multiple times in many cases, and their data have proven crucial in verifying this methodology and its results.

\begin{figure}
\includegraphics[width=\columnwidth]{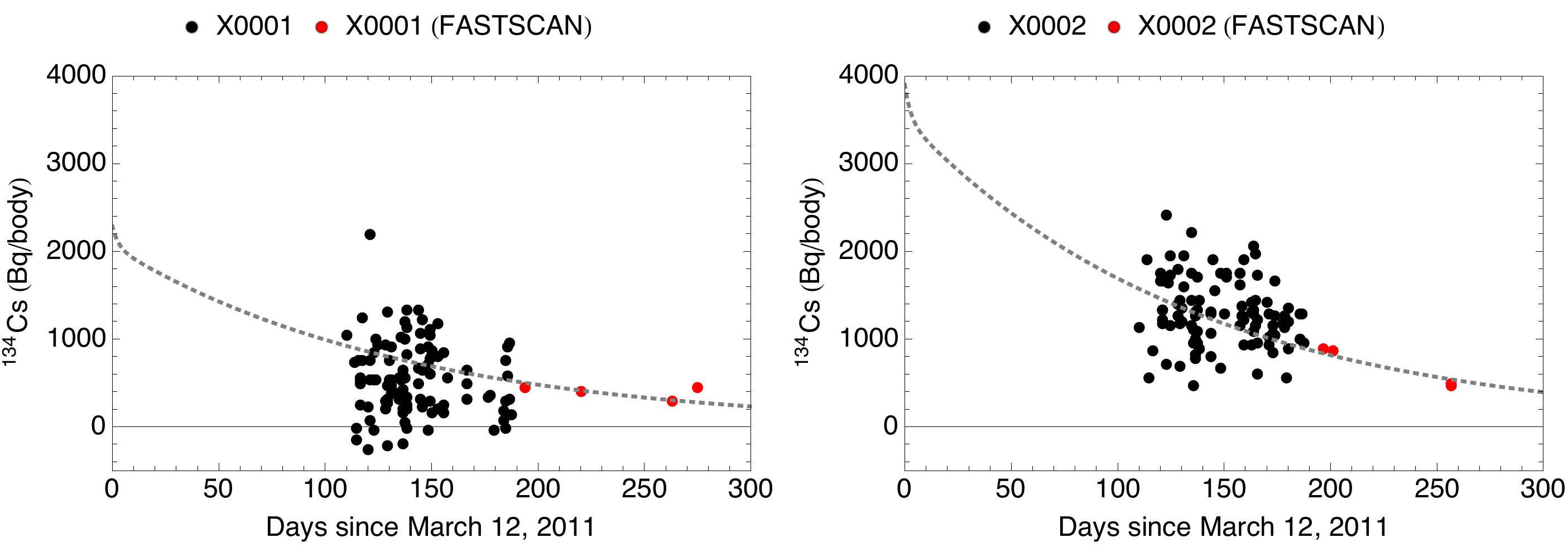}
\caption{\label{fig:x001x002} The $^{134}$Cs body burdens estimated in the present work (black dots) compared with those measured by FASTSCAN (red dots), for two MCGH staff members. The abscissa represents the number of days elapsed since March 12, 2011. The dotted curve in each panel is an exponential fit to the FASTSCAN data points using a $^{134}$Cs effective half life of 95 days.}
\end{figure}

Fig.~\ref{fig:x001x002} shows two such examples, male staff members identified as X0001 and X0002, both scanned more than 100 times with the Anzai chair-type WBC (black points) and more than 3 times with the FASTSCAN (red points).
The ordinate represents the amount of $^{134}$Cs (Bq/body), and the abscissa represents the number of days elapsed since March 12, 2011.
The dotted curve in each panel represents an exponential fit to the FASTSCAN data points assuming a $^{134}$Cs effective half life of 95 days~\cite{mondal3}\footnote{In 
Mondal 3\cite{mondal3}, the standard software used in Japan for calculating committed effective doses, $T_{1/2}^{\rm bio}=110$ days is used for the biological half life of radiocesium for adults. In the case of $^{134}$Cs, the effect of physical half life of $T_{1/2}^{\rm phi}=2$ years cannot be entirely ignored. We therefore use $(1/T_{1/2}^{\rm bio}+T_{1/2}^{\rm phi})^{-1} \approx 95$ days as the effective half life of $^{134}$Cs in the present analysis.}.

The data points from the Anzai WBC, after the compensation described above, are scattered, due to both the shorter measurement times and the lower detection efficiency, but they are consistent with the FASTSCAN results. 
We also note that additional radiocesium intake after July 2011 was negligibly small for both of these subjects~\cite{tsubo2}.

A consistency check of the data of less-frequently scanned hospital staff members is presented in Appendix \ref{sec:otherstaff}.

\subsection{Statistical and systematic uncertainties }

The statistical uncertainty of the WBC measurements was estimated from the data of 287 Group-A subjects, who were scanned twice on the same day. When the $^{134}$Cs count rate of a single measurement has a statistical uncertainty of $\sigma$, the uncertainty associated with the difference of two measurements must be $\sqrt{2} \sigma$. 

\begin{figure}
\includegraphics[width=0.6\columnwidth]{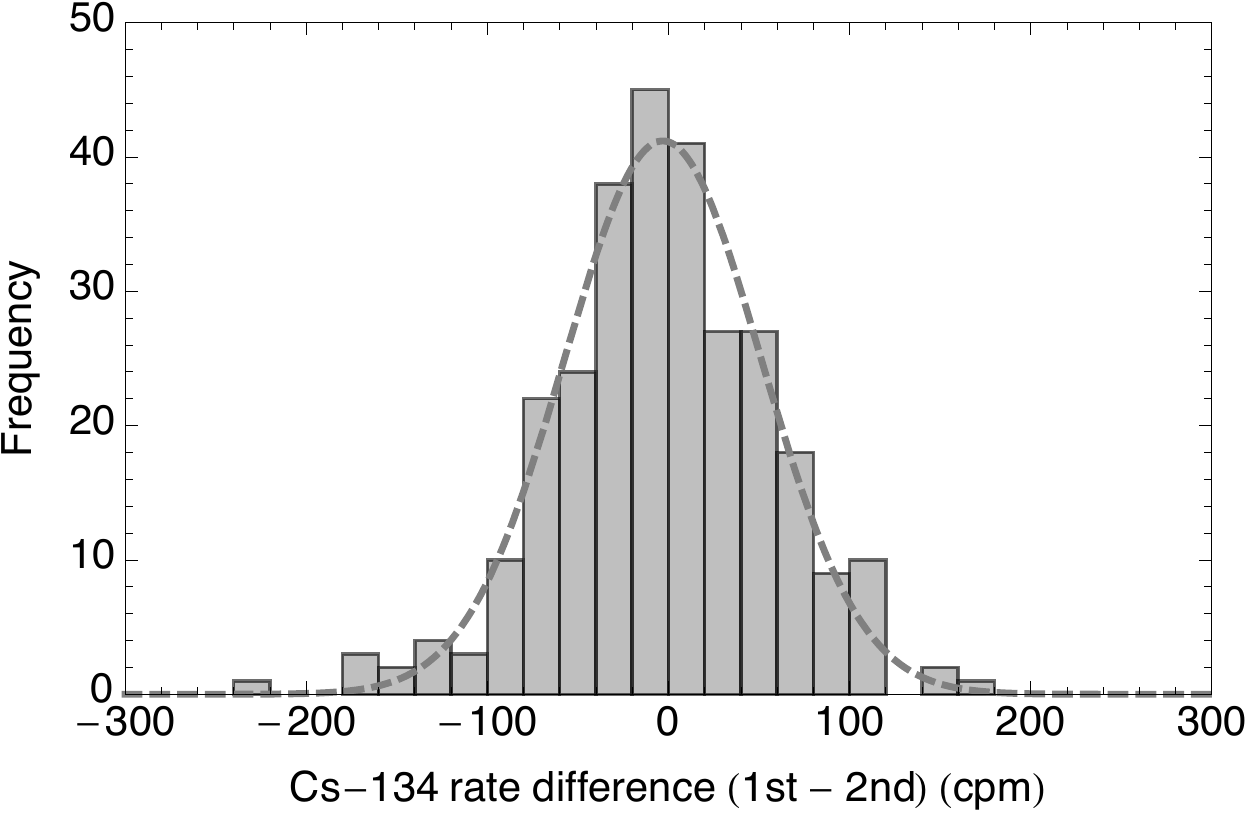}
\caption{\label{fig:difference} 
The difference in the $^{134}$Cs count rates (cpm) obtained during the first and second measurements conducted on the same day, for  287 subjects.}
\end{figure}

A Gaussian curve was fit to the the difference distribution shown in Fig.~\ref{fig:difference}, from which we obtained $\sqrt{2} \sigma = 54$ cpm, or $\sigma \sim 38$ cpm. With the efficiency coefficient described in Appendix \ref{sec:efficiency}, this corresponds to an uncertainty of $\sigma \sim 305$ Bq/body.

Systematic uncertainties can be inferred from the difference between the average of the last 10 measurements done with with the Anzai WBC in late August 2011, and the first FASTSCAN measurement, done in early September 2011, shown in Fig.~\ref{fig:x001x002}. For subject X0001 this was $\sim -110$  Bq, and for subject X0002 it was $\sim 110$ Bq (i.e., for X0001 the Anzai results are lower than the FASTSCAN, while for X0002 they are higher).
This $\sim 110 $Bq we take as an indicator of the uncertainty in the body-attenuated background estimation. Including the statistical fluctuation, the $1 \sigma$ uncertainty is estimated to be 300 Bq for each measurement, and there is an uncertainty of $\sim \pm 110$ Bq, common to all measurements, in the determination of the origin.% (i.e., the distribution shown in Fig.~\ref{fig:finaldistribution} can shift this degree either to the left or to the right).

\section{Results}

\begin{figure}
\begin{minipage}[t]{0.45\textwidth}
\includegraphics[width=\columnwidth]{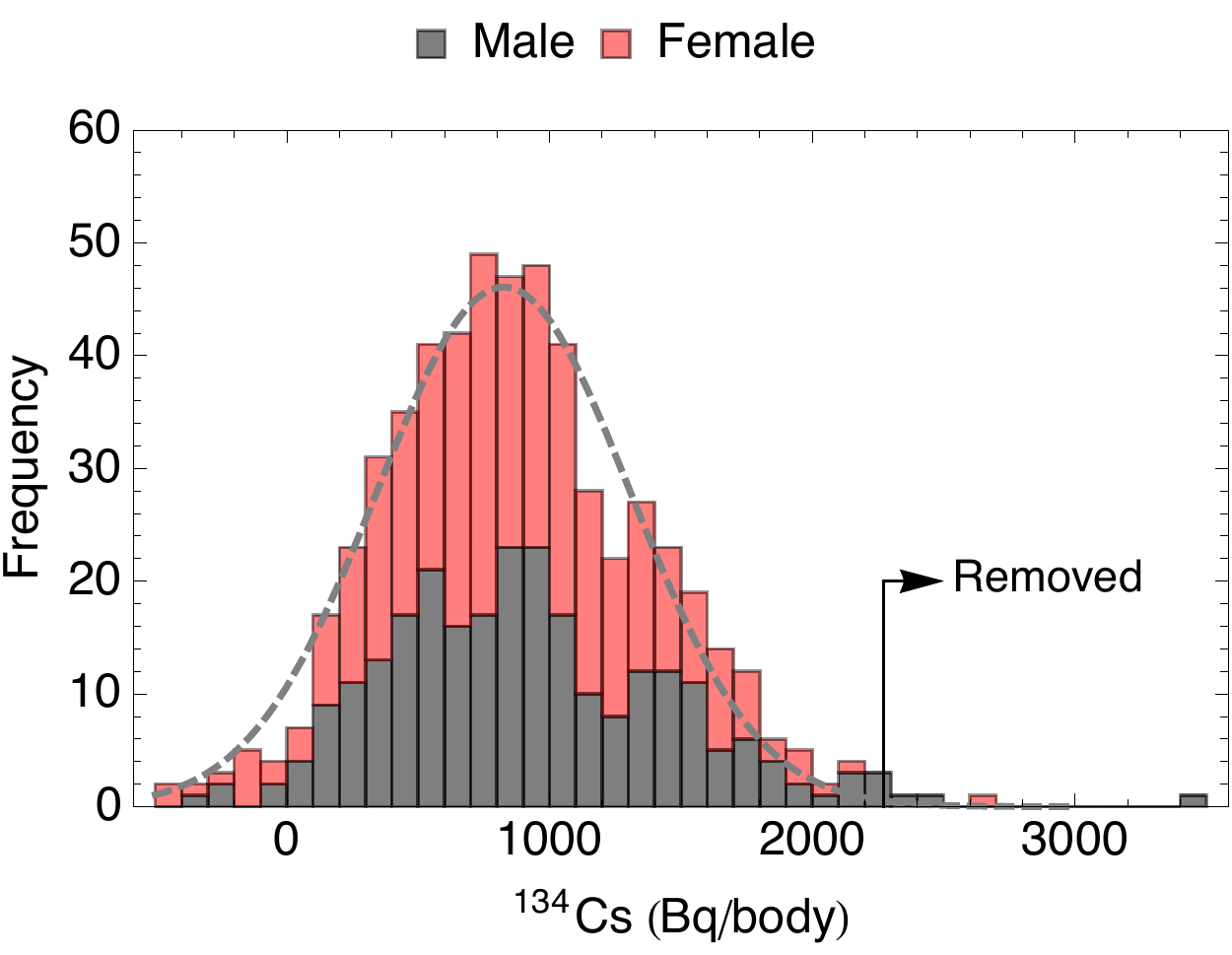}
\caption{\label{fig:finaldistribution} The estimated $^{134}$Cs body burden distributions of the Group-A male (black) and female (red, plotted above the male distribution), measured in July 2011, about 110-140 days after the Fukushima NPP accident. The gray dashed curve is a Gaussian fit to the sum of the male and female distributions, excluding the data points above 2,250 Bq.
}
\end{minipage}
\hfill
\begin{minipage}[t]{0.45\textwidth}
\includegraphics[width=\columnwidth]{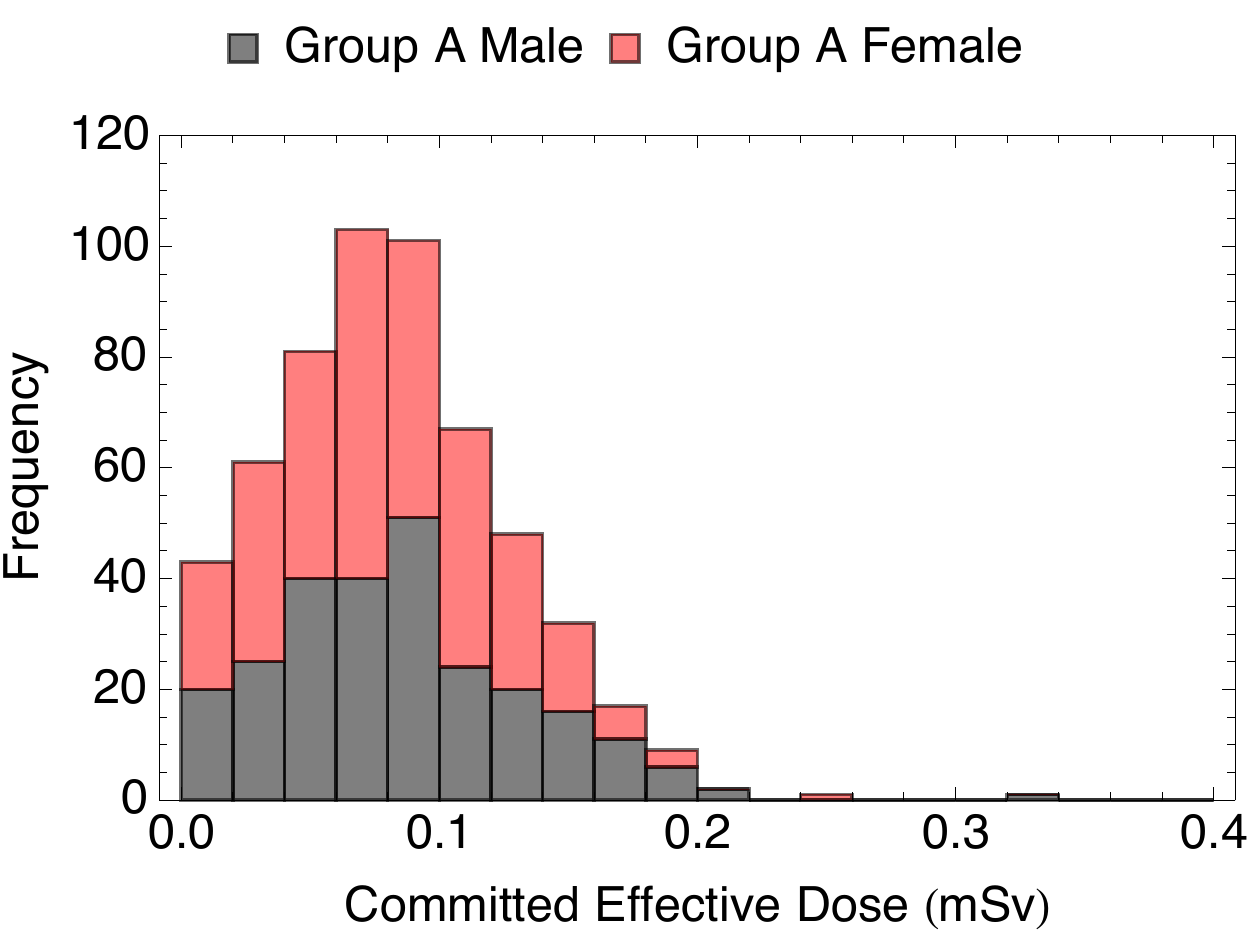}
\caption{\label{fig:fed} The estimated committed effective dose of the  Group-A male (black) and female (red, plotted above the male distribution).}
\end{minipage}
\end{figure}

The estimated $^{134}$Cs body burden distributions of the Group-A subjects is shown in Fig.~\ref{fig:finaldistribution}, as measured in July 2011. 
The distribution indicates that members of the high-risk group of Minamisoma residents surveyed all had less than 4000 Bq/body internal contamination from $^{134}$Cs on the day of the measurement.

As indicated by the dashed curve in Fig.~\ref{fig:finaldistribution}, the distribution is well represented by a Gaussian curve, except for 6 data points which lie above 2,250 Bq. 
The centroid of the distribution is 825 Bq, and the width $\sigma$ is 480 Bq. Subtracting the  uncertainty (statistical + background estimation) in quadrature, the average body burden of the Group-A subjects of Minamisoma is estimated to be $825 \pm 360 \pm 110$ Bq/body, where the last 110 is common to all subjects, which shifts the origin of the distribution shown in Fig.~\ref{fig:finaldistribution}.

We are able to estimate the distribution of the initial (March, 2011) $^{134}$Cs body burdens of this group 
assuming acute inhalation of radiocesium in March 2011, no further radiocesium intake until the day of the measurement, and using the $^{134}$Cs effective half life of 95 days.

From this distribution, and following the established $^{134}$Cs/$^{137}$Cs ratio of 1:1 in the early phase of the accident, the committed effective doses of all of these residents from these nuclides can be estimated to be $< 1$mSv as shown in Fig.~\ref{fig:fed}. This estimate does not include doses from $^{131}$I.

\section{Conclusions}
Using the first WBC unit installed in Fukushima Prefecture after the accident, the radiocesium body burdens of 566 high-risk residents of Minamisoma city were measured at the Minamisoma Municipal General Hospital. 
The analysis of the data was challenging because this chair-type WBC unit did not have sufficient shielding against background gamma rays, and methods had to be developed to reliably compensate for the body-attenuated background radiation.

Fortunately, data for repeated tests of MMGH staff members using both the Anzai chair-type and FASTSCAN WBC units were available, and could be used to check the validity of the analysis.

The CEDs of all subjects, estimated under the assumption of acute inhalation in March 2011, were found to be less than 1 mSv\footnote{ In retrospect, if the internal contamination of the subjects exceeded several tens of thousands Bq (i.e., CEDs of $\gtrsim$ several mSv), the $^{134/137}$Cs peak heights with the subject present would have been higher than those of the background alone, and it would have been possible to assess the doses with moderate precision, even without compensating for the body-attenuated background radiation.}.

During 2011, three other WBC surveys carried out:
\begin{itemize}
\item[i)] Matsuda et al.\ measured 173 evacuees and short-term visitors to Fukushima with the WBC at the Nagasaki University Medical School within one month after the Fukushima Dai-ichi accident,  and found the maximum CED to be 1~mSv~\cite{matsuda}. 

\item[ii)] Between June 27, 2011 and August 31 2011, almost the same time as the surveys at MMGH, residents of Kawamata, Namie and Iitate were measured with the WBC at the National Institute of Radiological Sciences (NIRS) in Chiba (174 subjects), and at the Japan Atomic Energy Agency (JAEA) in Ibaraki (3,199 subjects). The CEDs were found to be $< 1$ mSv for 99.8\% of the residents (5 subjects were between $1-2$~mSv, and 2 were between $2-3$~mSv)~\cite{nagataki,fukushimahp}. 
\item[iii)] Tsubokura et al.\ measured 9,498 Minamisoma residents with the FASTSCAN at MMGH between September 2011 and March 2012,  and found that the CEDs were less than 1 mSv in all but 1 resident (1.07 mSv)~\cite{jama}.
\end{itemize}
The present results, obtained with the first WBC unit installed in Fukushima Prefecture after the accident in Fukushima, are consistent with the results of other surveys. They all indicate that the early internal radiocesium doses of Fukushima residents were mostly $<1 $mSv.

\appendix
\section{Analysis of spectra\label{ap:spectra}}
Spectra similar to the one shown in Fig.~\ref{fig:spectra} were obtained for each individual test.  For each, the count rate of the $^{134}$Cs peak was obtained by fitting the following function with the chi-square method to the 796-keV peak, 
\begin{equation}\label{fitfunction}
p(E) =  \frac{A}{\sqrt{2 \pi \sigma^2}} e^{-\frac{(E - E_0)^2}{2 \sigma^2}} 
+ B \cdot {\rm erfc}(\frac{E - E_0}{\sigma}) + C,
\end{equation}
where the first term is the Gaussian peak; the second term, the complementary error function\footnote{The complementary error function is given by ${\rm erfc}(z)=1-{\rm erf}(z) = \frac{2}{\sqrt{\pi}} \int_0^z e^{-t^2} dt$.}, phenomenologically accounts for the contribution from scattered 796-keV gamma rays; and the last term represents a constant background.
The fit parameters are A: peak area, B: scattered gamma-ray contribution, C: constant background, $E_0$: peak centroid.  The peak width $\sigma$ was fixed by fitting a high-statistics background spectrum. A typical fit curve is shown in blue in Fig.~\ref{fig:spectra}.

We can similarly decompose the $^{134}$Cs--$^{137}$Cs double peak at 600-700 keV by adding multiple Gaussian-peak components to the fitting function. 
However, since the $^{137}$Cs peak count rates obtained by this kind of decomposition have larger statistical errors, hereafter only the $^{134}$Cs count rate, obtained as described above, is used. 

\section{A model for the body-attenuated background\label{ap:attenuate}}

The attenuation of the background radiation by the subject's body would depend on the body thickness, i.e. the sagittal dimension, if the background gamma rays are perpendicular to the chair back. We thus attempted to  parameterize the body-attenuated background level $BG$ to be:

\begin{equation}
BG(\alpha) = BG_0 (1 - \alpha \sqrt{w/h}),
\label{eq:eq2}
\end{equation}

\noindent
where $BG_0$  is the background level (measured without a subject), and 
 $\sqrt{w/h}$ is the body thickness parameter as defined by the ICRU  for mathematical phantoms~\cite{pcxmc},
 where $w$ represents body weight and $h$ represents body height\footnote{The use of a linearized form Eq.~\ref{eq:eq2} instead of an exponential form is justified because the range of $\sqrt{w/h}$ is small and $\alpha \sqrt{w/h}<1$.}.
 
\begin{figure}
\includegraphics[width=0.6\columnwidth]{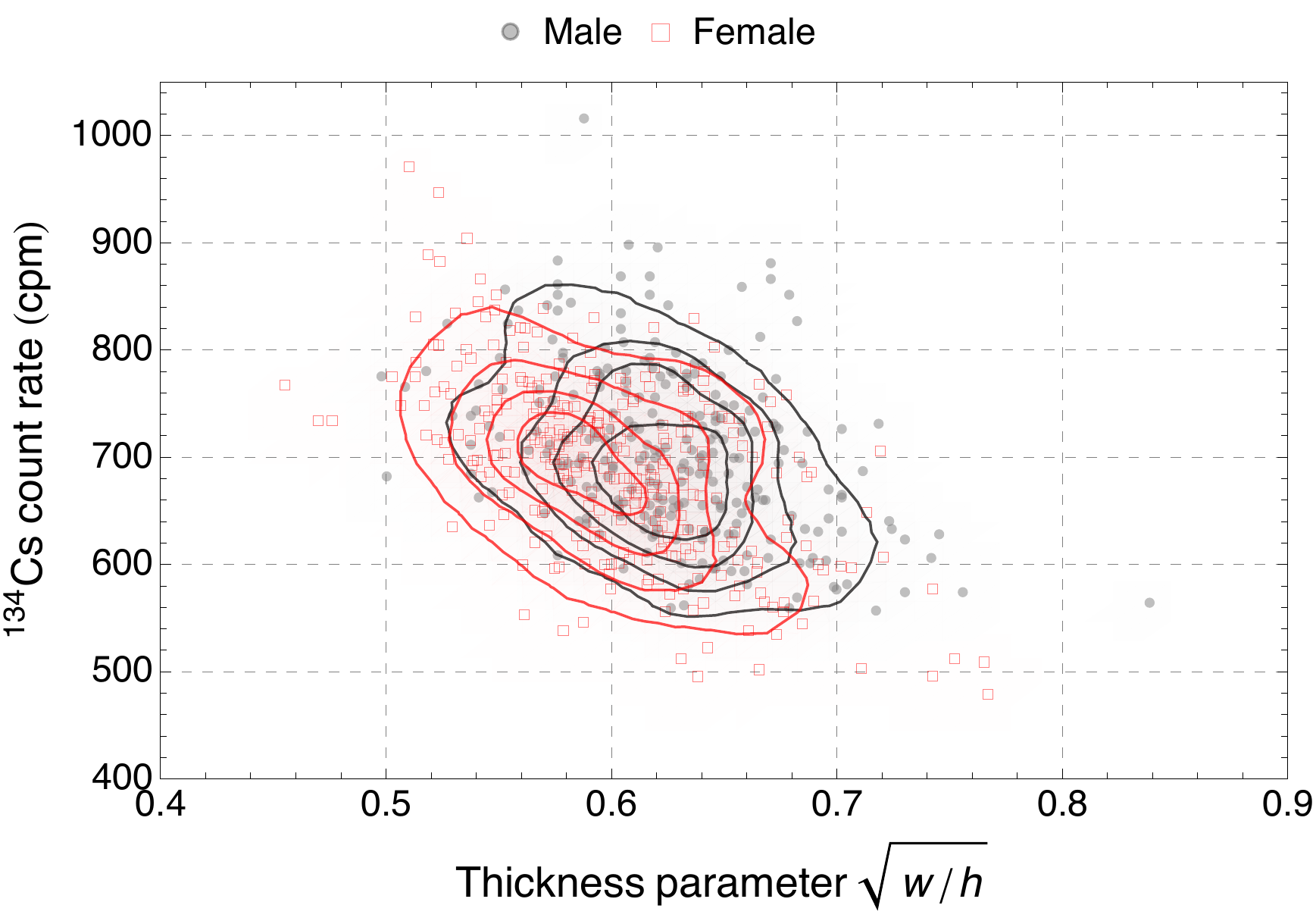}
\caption{\label{fig:thicknessvscs134} 
$^{134}$Cs count rates replotted against the body thickness parameter  $\sqrt{w/h}$, for male (black) and for female (red) subjects. }
\end{figure}

In Fig.~\ref{fig:thicknessvscs134}, we replot the same set of data shown in Fig.~\ref{fig:weightvscs134} against the body thickness parameter $\sqrt{w/h}$, together with 20-, 40-, 60- and 80-percentile contour lines.  

\begin{figure}
\includegraphics[width=0.6\columnwidth]{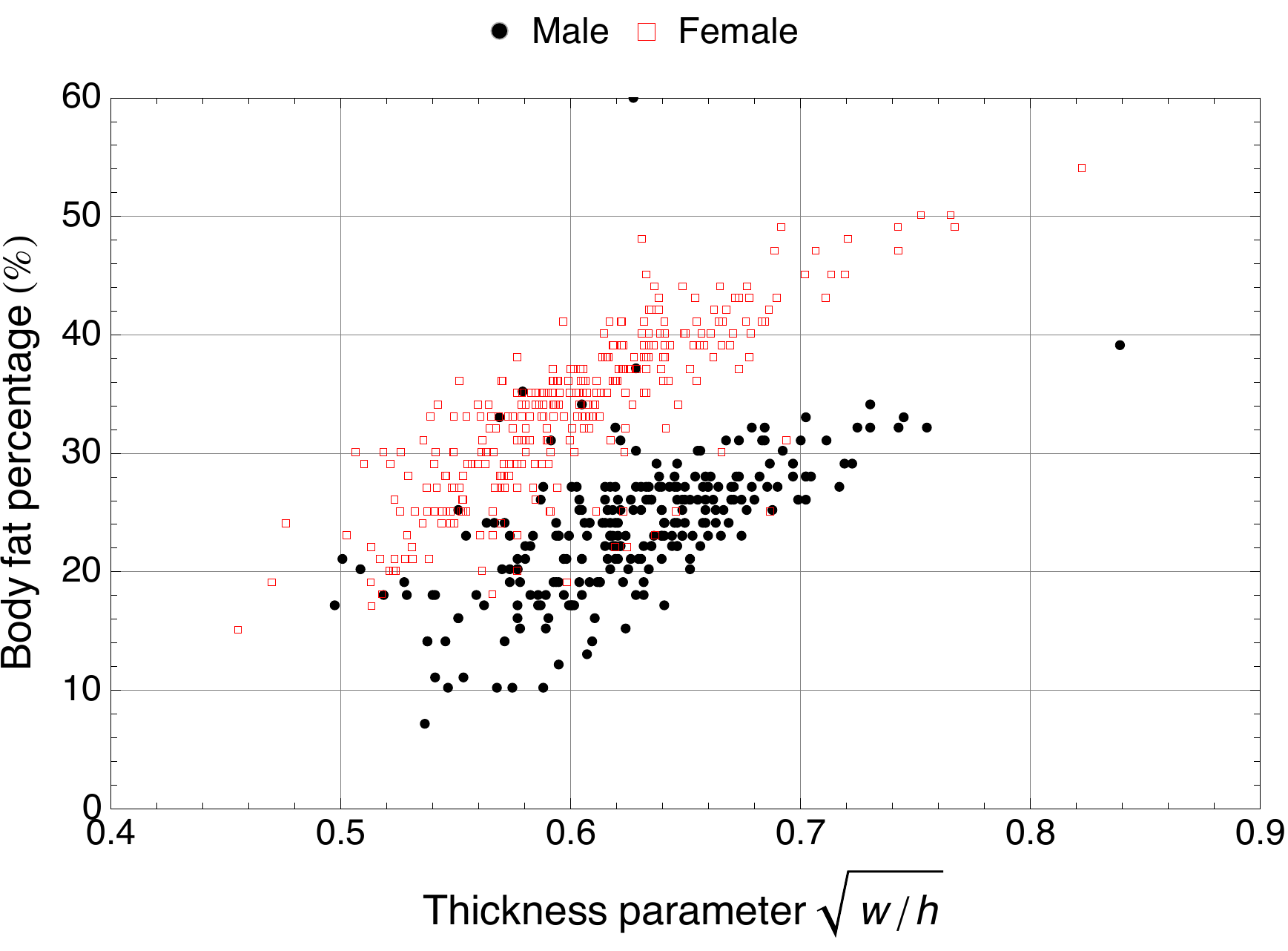}
\caption{\label{fig:bfp} The distribution of body thickness parameter $\sqrt{w/h}$ vs body fat percentage for male (filled circles) and female (red open squares) subjects.}
\end{figure}

Fig.~\ref{fig:thicknessvscs134} shows that the count rates for female subjects are lower than those for male subjects of the same $\sqrt{w/h}$. This we attribute to the difference in the gamma-ray attenuation coefficients between the fat and muscle tissues~\cite{tissue}, and in the difference of the body fat percentage between male and female subjects, depicted in Fig.~\ref{fig:bfp}.

We therefore modified Eq.~\ref{eq:eq2} to phenomenologically include the effect of the body fat percentage:

\begin{equation}
BG(\alpha, \beta) = BG_0 \left(1- \alpha (1+\beta \times {\rm BFP}) \sqrt{w/h}\right),
\label{eq:eq3}
\end{equation}

\noindent
where BFP is the body fat percentage. In this model, the {\em effective} body thickness parameter is now $(1+\beta \times {\rm BFP}) \sqrt{w/h}$.

\section{Detection efficiency\label{sec:efficiency}}

The detection efficiency coefficient of the WBC unit (peak counts per minute: cpm $\rightarrow$ Bq/body) was measured in the winter of 2011 by the manufacturer, using a phantom containing a combination of $^{134}$Cs and $^{137}$Cs sources, and was about 8 (Bq/body)/cpm. The efficiency varied from $\sim 7$ (Bq/body)/cpm to $\sim 9$ (Bq/body)/cpm for subjects' height between $120 \sim 190$ cm. This height dependence was taken into account in the analyses.

\section{Estimating the body burdens\label{ap:burden}}

If the body-attenuated background is properly subtracted, the body burden (Bq/body) should {\em not} depend on the body thickness nor the body fat percentage. The parameters $\alpha$ and $\beta$ of Eq.~\ref{eq:eq3} were determined so as to satisfy this requirement.

In detail, 1) for each subject, the net count rate was obtained by subtracting the estimated background $BG(\alpha,\beta)$; 2) the detection efficiency was multiplied to the net count to obtain the body burden (Bq/body); and 3) the {\em variance} of the body burden for all 566 data points was minimized by adjusting $\alpha$ and $\beta$. In this way, the dependence of the body burden on $\sqrt{w/h}$ and BFP was minimized.

The results are shown in Fig.~\ref{fig:contourswithbestparameters}, which depicts the correlation between the $^{134}$Cs body burden and: (left panel) the body thickness; (right panel) the body fat percentage. Neither of these distributions shows a clear dependence on $\sqrt{w/h}$  or on BFP, as intended. The difference between the male and female distributions found in Fig.~\ref{fig:thicknessvscs134} is no longer evident. The parameters obtained were $\alpha=0.34 { \pm 0.02 }$ and $\beta=1.06 {\pm 0.10}$
\footnote{Fig.~\ref{fig:contourswithbestparameters} includes six data points, shown as black diamonds, which lie distant from the main distribution. These data points were excluded when determining the $\alpha$ and $\beta$  parameters (as shown in Fig.~\ref{fig:finaldistribution}, these points are more than $3\sigma$ higher than the main body of data, which can be well represented by a Gaussian distribution).}.

\section{Data of other hospital staff members\label{sec:otherstaff}}
\begin{figure*}
\includegraphics[width=\textwidth]{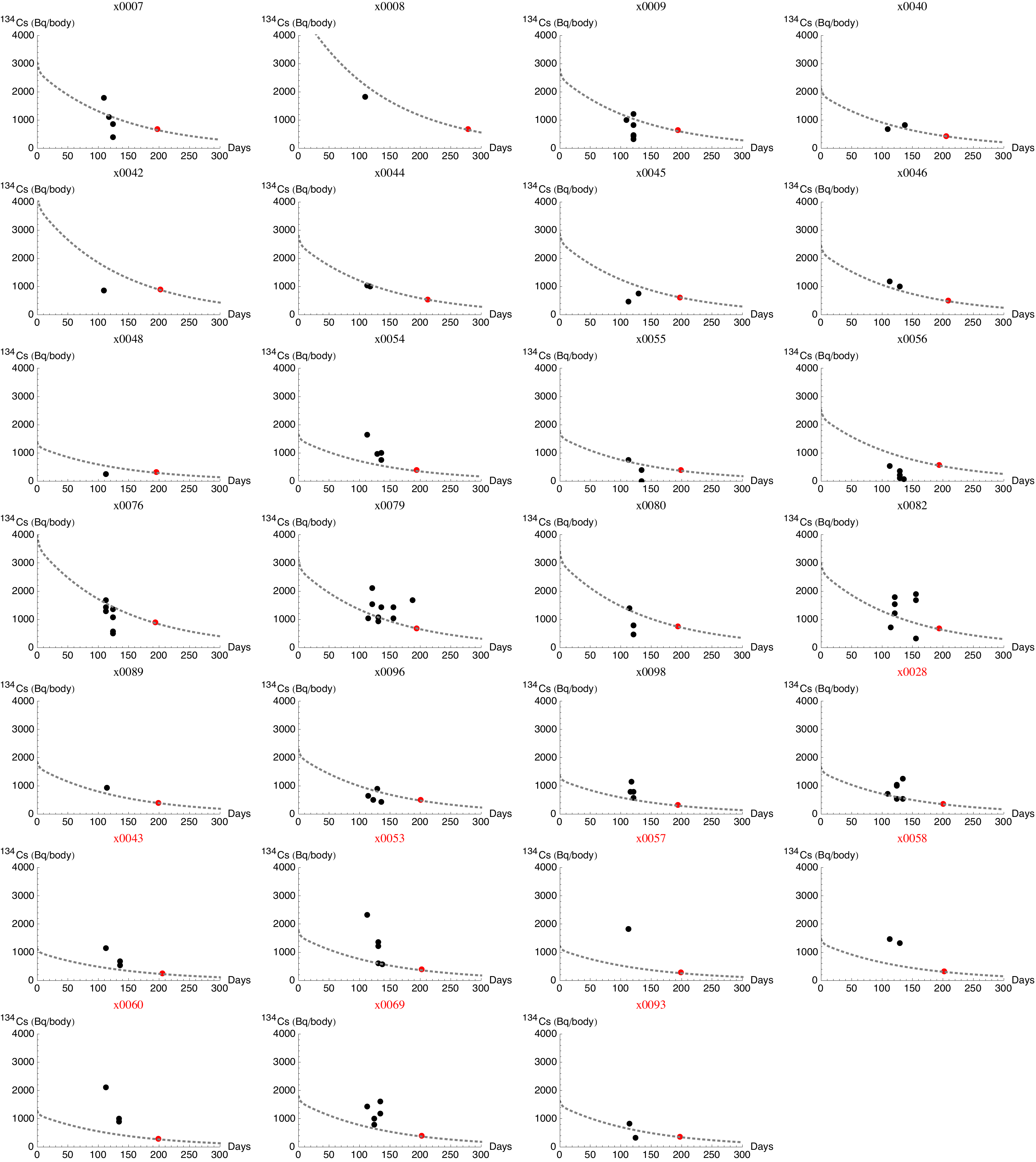}
\caption{\label{fig:xseries} As in Fig.~\ref{fig:x001x002}, but for other hospital staff members who were scanned less frequently. The first 19 panels are for male members and the last 8 are for female members.
}
\end{figure*}

Of the 101 other hospital staff members who were scanned less frequently, 37 were measured with both types of WBC, and 27 (19 males and 8 females) were found to be above the FASTSCAN detection limit. The results of those 27 members are shown in Fig.\ref{fig:xseries}.
Here again, the Anzai (black) and FASTSCAN (red) results are consistent within the margin of statistical and systematic errors. 
Unlike the cases of X0001 and X0002, these FASTSCAN measurements were performed typically $> 50$ days after the Anzai measurements, so it is not possible to directly compare the two sets. 

Here again, we note that additional radiocesium intake after July 2011 was negligibly small for all the MMGH hospital staff members, as the body burdens did not show noticeable increase.

\end{document}